\documentclass[11pt]{article}

\usepackage{color,xspace,wrapfig,fullpage}

\newcommand{\eg}{\emph{e.g.,}\xspace}
\newcommand{\Eg}{\emph{E.g.,}\xspace}

\usepackage{multirow,booktabs,amsmath,amsfonts,cite,hyperref,float,epsfig,graphics,stfloats,url,bm,booktabs,verbatim,pifont}
\usepackage[hyphenbreaks]{breakurl}
\newcommand{\cmark}{\ding{51}}
\newcommand{\xmark}{\ding{55}}

\begin{document}

\title{\LARGE \bf
An Energy Consumption Model for Electrical Vehicle Networks via Extended Federated-learning}
	
\author{Shiliang~Zhang\footnote{Computer Science and Engineering, Chalmers University of Technology. Email: shiliang@chalmers.se} }
	
\maketitle

\newcommand{\tabincell}[2]{\begin{tabular}{@{}#1@{}}#2\end{tabular}}

\begin{abstract}
Electrical vehicle (EV) raises to promote an eco-sustainable society. Nevertheless, the ``range anxiety'' of EV hinders its wider acceptance among customers. This paper proposes a novel solution to range anxiety based on a federated-learning model, which is capable of estimating battery consumption and providing energy-efficient route planning for vehicle networks. Specifically, the new approach extends the federated-learning structure with two components: anomaly detection and sharing policy. The first component identifies preventing factors in model learning, while the second component offers guidelines for information sharing amongst vehicle networks when the sharing is necessary to preserve learning efficiency. The two components collaborate to enhance learning robustness against data heterogeneities in networks. Numerical experiments are conducted, and the results show that compared with considered solutions, the proposed approach could provide higher accuracy of battery-consumption estimation for vehicles under heterogeneous data distributions, without increasing the time complexity or transmitting raw data among vehicle networks.
\end{abstract}

\section{Introduction}

Electric vehicle (EV) has been rolling out as an enormous contributor to the eco-stainable industry. Nevertheless, through years of penetration, EV has not significantly improved its attractiveness among customers~\cite{AcceptanceSurvey1,AcceptanceSurvey2}. The long-existed ``range anxiety''~\cite{RangeAnxiety3} induced by limited battery volume and lengthy charging time hinders EV's broader acceptance~\cite{HurdlesForEV}. Such disadvantages have motivated the rise of energy-efficient solutions for EVs to decrease range anxiety. In this paper, we analyze existing energy-efficient approaches, and try to come up with a novel solution that fits the need of EV networks better.

Energy-efficient route planning, a.k.a e-route~\cite{e-route-recently}, has been proposed to reduce battery usage by providing route options with less energy consumption, aiming to benefit sustainable range, reduce charging frequency, and lessen range anxiety. The key objective of e-route is to derive a model that estimates EV energy consumption for a specific route. Investigations on such models have been conducted considering various factors. Most approaches take inputs as road gradient and route length~\cite{eRout1,eRout2,eRout3,eRout4,eRout5,eRout6,eRout7} in energy estimation, and other factors like vehicle specifications and driving behavior are also considered~\cite{eRout1,eRout6,eRout5}. With input factors determined, a model can be established via recording all the considered factors in a testing driving trip, followed by training and generating the model parameters based on the recorded data~\cite{eRout7,eRout8}.

Existing e-route approaches often include a great variety of input factors, such as the specifications of the EV's power train, heater and air-conditioner, wiper, radio~\cite{eRout6}, atmosphere humidity, traffic light~\cite{eRout5}, and even driver's weight~\cite{eRout7,eRout13}. The vehicle's driving and stationary status are also considered in~\cite{DrivingAndStationary}. The authors in~\cite{InputFactorsMore} develop a model with more inclusive inputs, including air density, rolling resistance, vehicle frontal area, wind speed, road surface angle, vehicle acceleration, and vehicle operating condition. A Bellman-Ford-based e-route method in~\cite{BellmanFordBasedMethod} is presented with input factors such as precipitation, light, temperature, crossing conditions, and road type. The underlying motivation to pursue more input features is to provide an inclusive model that fits for vehicles with diversities of specifications~\cite{eRout8} or vehicles with heterogeneous data distributions. \Eg an estimation model that covers the need of vehicles under city-street driving, highway driving, mountain driving, etc, where more possible input factors are included to gain various data populations with higher granularity. Nevertheless, such a single inclusive model can bring obstacles to the model training: it is not clear how to collect the needed datasets for covering all essential ranges of diversities of vehicle populations. Moreover, learning forcedly on the aggregation of heterogeneous data distributions might entail a model that diverges all data distributions~\cite{FederatedLearningOnIIDdata}. Such approaches are also vulnerable to abnormal vehicle behaviors, \eg erratic driving patterns, which generate data that can severely degrade any learned model.

Federated-learning~\cite{FederatedLearning} (FL) provides a potential solution to heterogeneous data distribution in network learning, which leverages shared information among network for cooperative learning that avoids sending raw data~\cite{Federated-learningAndPrivacyProtection}. Some recent FL approaches have adopted advanced features that, distinctions between different populations of network members are considered in model learning. We review the existing FL solutions to the challenge of heterogeneous data distribution in the following.

\subsection{Related works}

\subsubsection{Sharing small amounts of raw data}

This approach circumvents data distributions' diversity via sharing a small percentage of local data from each network participant~\cite{FederatedLearningOnIIDdata,FLChallenge_nonIID2}. During each learning iteration, all the participants transmit a small proportion of their local data to the learning server, where all the shared data are integrated to establish the model. The size of the shared data can be considerably large as the network size grows, resulting in an inclusive dataset for the learning server that covers various vehicle populations. While this approach reduces data heterogeneity by an inclusive data collection, directly sharing raw data raises concerns regarding privacy-sensitive information since it breaks the key guidelines for the framework of federated-learning.

\subsubsection{Assuming a statistical relationship among local populations}\label{section: assuming relationship between network members}

This approach makes assumptions on the relationship between network members. \Eg the work in~\cite{FLChallenge_MTL} assumes the covariance amongst learned results from different network members exists, and optimizes model parameters based on such covariance. The optimization of this approach requires transmitting raw data to the optimizing engine, which poses privacy vulnerabilities. A neural network model with shared parameters in lower layers is assumed in~\cite{FLChallengs_NNModel}, where a global objective function is optimized to acquire the desirable model parameters. However, those assumptions might not fit practical scenarios in which these relationships are not necessarily known to exist.

\subsubsection{Assuming on the classification of data populations}
The third approach classifies network members into several groups named by, \eg blocks~\cite{FLChallenge_block} or clusters~\cite{FLChallenge_Clustering,FlChallenge_Clustering2}, where members belong to the same group are regarded as following the same distribution. Then the model learning task is converted into optimizing model parameters within each group, so that one vehicle can avoid learning on data generated from a different vehicle population. The performance of this approach will depend significantly on the grouping efficiency. However, this approach does not apply to small scale networks, where no efficient groups could be extracted.

\subsubsection{Restricting measures on local learned results}\label{section: restricting learned results}
This method exploits more features of the learned results, \eg a new objective function for federated-learning is proposed in~\cite{FLChallenge_LimitedLocalImpact} to limit the impact of learned results on model integration, aiming to enhance convergence in the case of heterogeneous data distributions. A similar modeling method can be found in~\cite{FLChallenge_LimitLearnedResultImpact2} to improve robustness against heterogeneity by adding a penalty on the norm of learned results in the objective function. Nevertheless, this approach slows down the influence of heterogeneity rather than eliminating it.

Beyond all the pros and cons, those mentioned methods are vulnerable to anomalies. Abnormal data, \eg data generated by vehicles under erratic driving behaviors or polluted by severe noise, can be propagated into the learning engine and degrade any learned model.

\subsection{Our contribution}

We provide an important solution for e-route with an estimation model based on federated-learning, which could preserve energy-estimation accuracy in the presence of data distribution heterogeneities in EV networks. The proposed approach, to the best of our knowledge, is the first to combine federated-learning with EV energy-consumption estimation. The new approach extends the federated-learning structure with two components: (i) anomaly detection and (ii) sharing policy. The former component keeps network learning free from preventing factors like data from erratic-driving vehicles, while the latter guarantees that during network learning, one vehicle will not learn from vehicles behaving significantly different from it when vehicles of different populations exist in the same network. Beyond that, our approach differs from existing ones with no assumed prior knowledge on data, nor transmitting raw data in model learning. We conduct numerical experiments based on real-world data to verify the model prediction performance under heterogeneous data distributions.

This paper's organization is as follows: Section~\ref{section: algorithm description} describes the proposed approach named by extended federated-learning in detail; Experiments and results on real-world data are shown in section~\ref{section: experiments}; Conclusions are given in section~\ref{section: conclustion}.

\section{Methodology: Extended Federated-learning}\label{section: algorithm description}

\begin{figure}[hbp]
  \centering
  \includegraphics[width=0.9\textwidth]{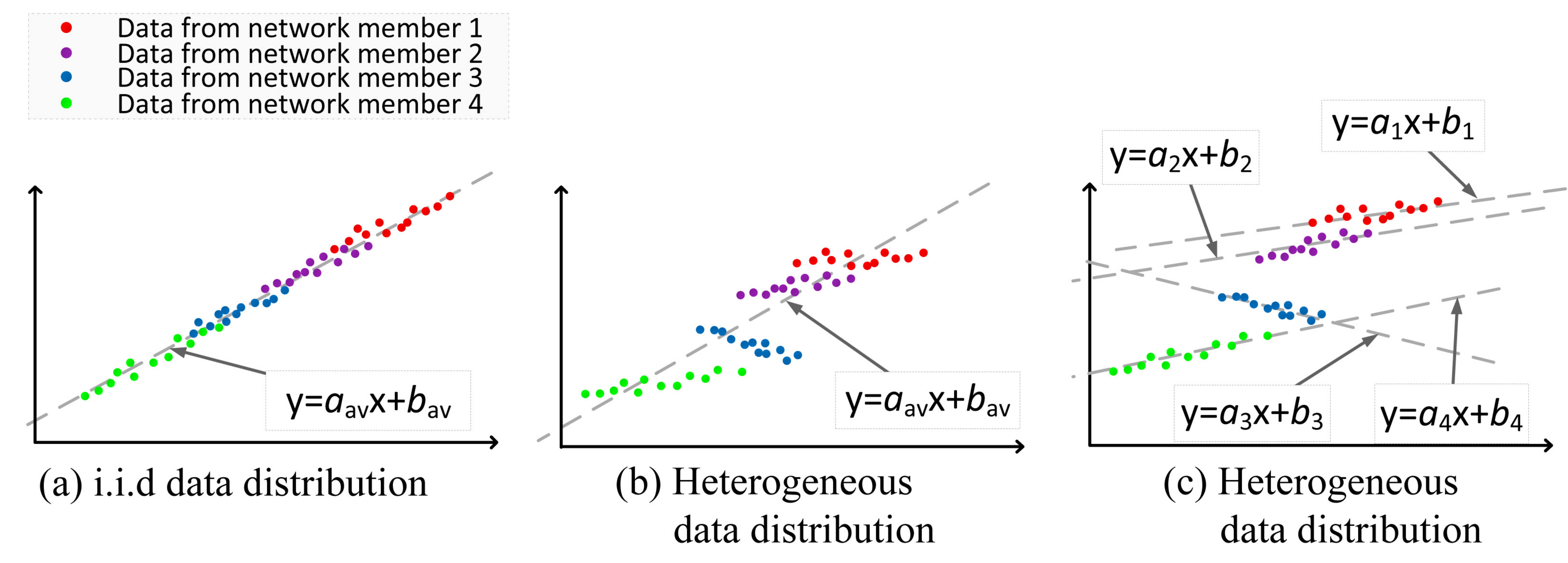}\\
  \caption{\label{fig: data distributions}Federated-learning on uniformed and heterogeneous data distribution}
\end{figure}

This section illustrates how the proposed extended federated-learning is constructed, and how its two components of anomaly detection and sharing policy collaborate to address the data distribution heterogeneity issue.

We first give examples of learning scenarios on the identical independent distribution (i.i.d) and heterogeneous data distribution in Figure~\ref{fig: data distributions}. Though generated by different vehicles, the data in Figure~\ref{fig: data distributions}-(a) follows i.i.d since they are likely to be generated by the same mechanism, and data from each member benefits the same model. Such dataset fits well in approaches that integrate all information to generate a final model, \eg $y=a_{av}x+b_{av}$ for every member. However, it is often heterogeneous rather than i.i.d data distribution that is available in practical scenarios, as shown in Figure~\ref{fig: data distributions}-(b), where the data is generated from different types of vehicle trips such as city-street, highway, mountainous road, etc. Those data under heterogeneous distribution are hard to be covered by one single model, as shown in Figure~\ref{fig: data distributions}-(b), where one single model like $y=a_{av}x+b_{av}$ is inaccurate to make predictions for all the data populations. Beyond this data heterogeneity issue, data generated abnormally, \eg data from vehicles under erratic drivers, can lead to the uselessness of any learned models.

\begin{figure}[!htbp]
  \centering
  \includegraphics[width=0.65\textwidth]{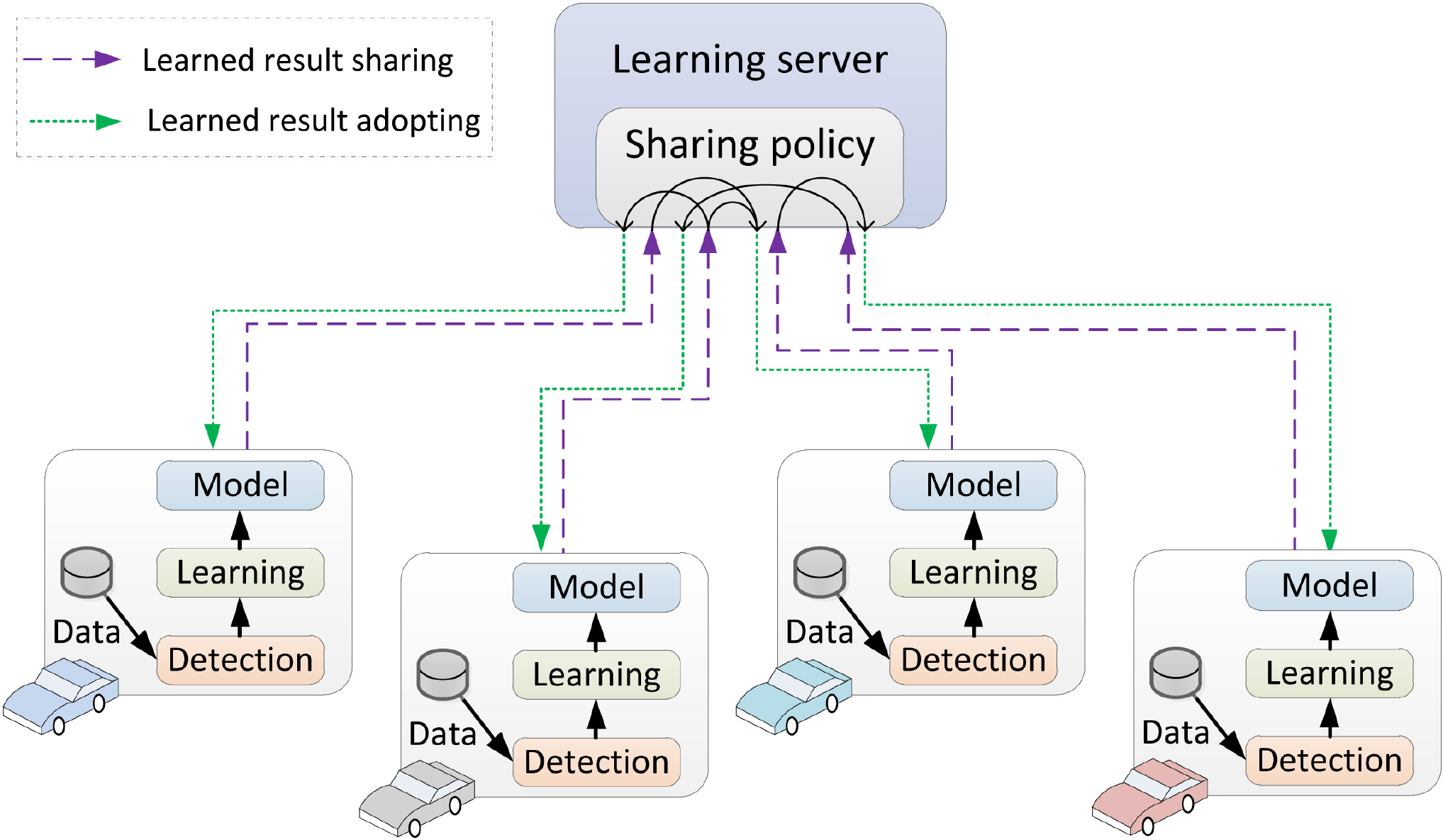}\\
  \caption{\label{fig: proposed model}The proposed model structure}
\end{figure}

This section depicts a network learning approach that assigns different models to different network participants, as shown in Figure~\ref{fig: data distributions}-(c), and the learned result from each vehicle can be shared and leveraged to enhance other vehicle's learning. The rule for sharing of learned results is determined by the developed sharing policy. Before any of the data are leveraged in learning procedures, anomaly detection is performed to filter out abnormal data. Figure~\ref{fig: proposed model} illustrates the proposed extended federated-learning, which includes the two developed components of anomaly detection and sharing policy.

The two components collaborate as follows. The anomaly detection filters out abnormal data and provides informative data for model learning, and also informs the sharing policy that the abnormal vehicles need to extract learned results from other vehicles, so that the vehicle's learning procedure stuck by anomalies can be continued. The sharing policy checks the available learned results shared from anomaly-free vehicles, selects those beneficial for the abnormal vehicle, and transmits the selected results to the vehicle in need. As for the anomaly-free vehicles, they learn models based on their locally generated data. During this collaboration, the anomaly detection determines when to conduct sharing, while the sharing policy makes sure that the shared content can benefit the receiving vehicles. As a result, in the case of heterogeneous data distribution, vehicles of different populations learn models either from their local data, or via extracting other vehicles' learned results capable of preserving the learning procedures of the considered vehicle.

We present the developed scheme via algorithmic pseudo-code in Figure~\ref{fig: pseudo-code} for more technical presentation, where we adopt stochastic gradient descent as the learning engine and the gradient in learning procedure as the learned result. In the following content, we will illustrate in detail how the anomaly detection and sharing policy are constructed and executed in detail.

\begin{figure}[!htp]
\begin{minipage}{\linewidth}
\centering
\small
\begin{tabular}{p{0.96\textwidth}}
\toprule
\textbf{Initial settings:} \\
number of network members: $N$;\\
threshold for model update: $\delta$; \\
anomaly detection settings: $Anomaly_{(\theta,K)}(\,)$; \\
sharing policy settings: $share_{\mu}(\,)$;\\
empty set for abnormal data: $abnor\{\,\}$; \\
empty set for normal data: $normal$\{\,\};\\
\textbf{Inputs:}\\
input-output data of the $i$-th member at the $k$-th iteration: \{$\bm{x}_{i}^{k},y_{i}^{k}$\}, $i\in\{1,2,3,...,N\}$;\\
the $i$-th local gradient at the $(k-1)$-th iteration: $\bm{g}_{i}^{k-1}$, $i\in\{1,2,3,...,N\}$; \\
the $i$-th local model parameter at the $(k-1)$-th iteration: $\bm{\omega}_{i}^{k-1}$, $i\in\{1,2,3,...,N\}$; \\
\textbf{Calculations for the $k$-th iteration:}\\
\textbf{for} $i\in\{1,2,3,...,N\}$\\
\qquad conduct anomaly detection on the $i$-th data via\\
\qquad $Anomaly_{(\theta,K)}(\bm{x}_{i}^{k},y_{i}^{k})$\\
\qquad \textbf{if} \{$\bm{x}_{i}^{k},y_{i}^{k}$\} is an anomaly:\\
\qquad \qquad add the member index $i$ into $abnor\{\,\}$;\\
\qquad \textbf{else}\\
\qquad \qquad add the member index $i$ into $normal$\{\,\};\\
\qquad \qquad calculate $\varepsilon=\bm{x}_{i}^{k}\cdot\bm{\omega}_{i}^{k-1}-y_{i}^{k}$;\\
\qquad \qquad \textbf{if} $\|\varepsilon\| \le \delta$:\\
\qquad \qquad \qquad no need to update;\\
\qquad \qquad \qquad $\bm{\omega}_{i}^{k}=\bm{\omega}_{i}^{k-1}$;\\
\qquad \qquad \textbf{else}\\
\qquad \qquad \qquad update the local model via stochastic gradient descent~\cite{StochasticGradientDecent}:\\
\qquad \qquad \qquad $\bm{g}_{i}^{k}=\frac{\partial(y_{i}^{k}-\bm{x}_{i}^{k}\cdot\bm{\omega}_{i}^{k-1})^{2}}{\partial \bm{\omega}}\cdot \lambda$;\\
\qquad \qquad \qquad $\bm{\omega}_{i}^{k}=\bm{\omega}_{i}^{k-1}+\bm{g}_{i}^{k}$;\\
\qquad \qquad \textbf{end}\\
\qquad \textbf{end}\\
\textbf{end}\\
\textbf{for} $j\in abnor\{\,\}$\\
\qquad find all the gradients $\bm{g}_{*}^{k} \; where\; *\in normal\{\,\}$ that $share_{\mu}(\bm{g}_{j}^{k-1},\bm{g}_{*}^{k})\le\mu$;\\
\qquad take the average of the found gradients, denoted by $\bm{\overline{g}}_{*}^{k}$;\\
\qquad update the $j$-th local model as:\\
\qquad $\bm{g}_{j}^{k}=\bm{\overline{g}}_{*}^{k}$;\\
\qquad $\bm{\omega}_{j}^{k}=\bm{\omega}_{j}^{k-1}+\bm{\overline{g}}_{*}^{k}$;\\
\qquad \textbf{if} no such gradient is found:\\
\qquad \qquad keep the $j$-th local model parameter unchanged:\\
\qquad \qquad $\bm{\omega}_{j}^{k}=\bm{\omega}_{j}^{k-1}$;\\
\qquad \textbf{end}\\
\textbf{end}\\
the $k$-th iteration ends, and move towards the next iteration\\
\bottomrule
\end{tabular}
\end{minipage}
\caption{\label{fig: pseudo-code}Pseudo-code for learning procedures of the developed extended federated-learning model}
\end{figure}

\subsection{Anomaly detection}

We refer to the term ``anomaly'' as preventing factors against learning convergence in a dataset. Such anomaly, \eg vehicle under erratic driving behaviors, brings data that can render any model to be useless and should be filtered out in model learning. However, there is still an absence of anomaly-free data for learning after the removal of abnormal data. Fortunately, this absence can be filled in network learning via exploiting learned results from anomaly-free vehicles of the same network.

The proposed approach carries out anomaly detection upon each vehicle's local data based on sequential statistical features, and anomalies will be detected and prevented from contributing data to the learning engine. Furthermore, to avoid interruption on learning posed by abnormal data, the vehicles with anomaly will try to adopt learned results from the anomaly-free vehicles, with details shown in Section~\ref{subsection: sharing policy}. In this section, we describe how the anomaly detection mechanism is constructed.

Suppose the local dataset for an individual vehicle at the $n$-th instant is accumulated as $\{(\bm{x}_{1},y_{1}),(\bm{x}_{2},y_{2}),$ ... $,(\bm{x}_{i},y_{i}),$ ... $,(\bm{x}_{n},y_{n})\}$, where $\bm{x}_{i}$ and $y_{i}$ are the input and output data at the $i$-th instant, respectively, and $\bm{x}_{i}=\{x_{i}^{1},x_{i}^{2},...,x_{i}^{m}\}$, $m$ denotes the number of input attributes. When the data at the $(n+1)$-th instant comes as $(\bm{x}_{n+1},y_{n+1})$, we want to detect whether the newcomer is an anomaly or not.

The detection result will be decided based on the $K$ most recent entries of the dataset, where $K$ is a pre-defined coefficient. Those $K$ entries will be used to generate a predictive region $\bm{\gamma}$ where the $(n+1)$-th input data $\bm{x}_{n+1}$ is supposed to be located. The predictive region $\bm{\gamma}$ is constructed based on the expectation $\bm{E}_{K}$ and the standard derivation $\bm{\sigma}_{K}$ of the $K$ entries as:

\begin{equation}\label{equation:predictive region}
\begin{split}
\bm{\gamma}=[\bm{E}_{K}-\theta\bm{\sigma}_{K},\bm{E}_{K}+\theta\bm{\sigma}_{K}]\\
\end{split}
\end{equation}
where $\theta$ is a pre-defined coefficient and

\begin{equation}\label{equation:expectation and standard deviation}
\begin{split}
\bm{E}_{K}=\frac{1}{K}\sum\limits_{i=n-K+1}^{n}\bm{x}_{i}\\
\\
\bm{\sigma}_{K}=\sqrt{\frac{1}{K}\sum\limits_{i=n-K+1}^{n}(\bm{x}_{i}-\bm{E}_{K})^{2}}\\
\end{split}
\end{equation}
notice that $\bm{\gamma}$, $\bm{E}_{K}$, and $\bm{\sigma}_{K}$ are with the same dimension of $1\times m$. The lower and upper border of $\bm{\gamma}$ can be presented as:

\begin{equation}\label{equation:upper and lower border}
\begin{split}
\bm{\gamma}_{l}=\bm{E}_{K}-\theta\bm{\sigma}_{K}\\
\\
\bm{\gamma}_{u}=\bm{E}_{K}+\theta\bm{\sigma}_{K}\\\\
\end{split}
\end{equation}
and then with the predictive region established, the anomaly detection operates as follows: for any of the element $x_{n+1}^{j}, j\in\{1,2,...,m\}$ in $\bm{x}_{n+1}$, if they all satisfy that:

\begin{equation}\label{equation:anomaly detection}
\begin{split}
\gamma_{l}^{j}\le x_{n+1}^{j}\le\gamma_{u}^{j}\\
\end{split}
\end{equation}
where $\gamma_{l}^{j}$ and $\gamma_{u}^{j}$ are the $j$-th element of $\bm{\gamma}_{l}$ and $\bm{\gamma}_{u}$, respectively, then the input data at the $(n+1)$-th instant $\bm{x}_{n+1}$ is regarded as anomaly-free. Otherwise, $\bm{x}_{n+1}$ will be treated as anomaly data and filtered out of learning procedures.

This anomaly detection mechanism guarantees that abrupt change in the data stream could be captured and filtered to preserve learning convergence. The pre-defined coefficients $K$ and $\theta$ could provide adjustable detection sensitivity and tolerance toward abrupt changes, respectively.

In the following section, we present the principle in extracting learned results from the other vehicles when one suffers anomalies.

\subsection{Sharing policy}\label{subsection: sharing policy}
Suppose $\alpha$ and $\beta$ are two vehicles in the same network with data entry $(\bm{x}_{n^{\alpha}},y_{n^{\alpha}})$ and $(\bm{x}_{n^{\beta}},y_{n^{\beta}})$ at the $n$-th instant, respectively, and the data entry from $\alpha$ is detected as abnormal while that of $\beta$ is anomaly-free. With this detection result, $\alpha$'s current model parameter $\bm{\omega}_{n-1}^{\alpha}$ cannot be renewed by $(\bm{x}_{n^{\alpha}},y_{n^{\alpha}})$, and $\beta$ can successfully learn from $(\bm{x}_{n^{\beta}},y_{n^{\beta}})$ to obtain its learned result $\bm{\iota}_{n}^{\beta}$ and update its model parameter via method \eg stochastic gradient descent as:

\begin{equation}\label{equation:model parameter renew for Beta}
\begin{split}
\bm{\iota}_{n}^{\beta}=\frac{\partial(y_{n^{\beta}}-\bm{x}_{n^{\beta}}\cdot\bm{\omega}_{n-1}^{\beta})^{2}}{\partial \bm{\omega}_{n-1}^{\beta}}\cdot \lambda\\
\\
\bm{\omega}_{n}^{\beta}=\bm{\omega}_{n-1}^{\beta}+\bm{\iota}_{n}^{\beta}
\end{split}
\end{equation}
where $\bm{\omega}_{n-1}^{\beta}$ is $\beta$'s current model parameter and $\bm{\omega}_{n}^{\beta}$ is the renewed one, $\lambda$ is a predefined learning step. To preserve the learning procedures of $\alpha$, it is intuitive to try to introduce $\beta$'s learned result $\bm{\iota}_{n}^{\beta}$ to $\alpha$ to enable the continuity of $\alpha$'s learning procedure, since $\alpha$ and $\beta$ are in the same network and potential of enhancing each other's learning performance. However, there is the question that whether $\bm{\iota}_{n}^{\beta}$ can benefit $\alpha$'s learning iterations. To answer this question and share learned results with higher confidence, we derive the sharing policy in this section.

Our hypothesis for confident and efficient sharing is that the learned result to be shared shows consistency with the receiving vehicle's learning procedures, and such consistency is measured by \emph{similarity} in this paper. Take the example of $\alpha$ and $\beta$, and we define the \emph{similarity} $\rho$ between $\alpha$'s most recent available learned result $\bm{\iota}_{n-1}^{\alpha}$ and $\beta$'s current learned result $\bm{\iota}_{n}^{\beta}$ as:

\begin{equation}\label{equation:defination of similarity}
\begin{split}
\rho=1-\left| \frac{\bm{\iota}_{n-1}^{\alpha}}{\lvert \bm{\iota}_{n-1}^{\alpha}\rvert}-\frac{\bm{\iota}_{n}^{\beta}}{\lvert \bm{\iota}_{n}^{\beta}\rvert} \right|\\
\end{split}
\end{equation}
where higher $\rho$ indicates more similarity in vector direction between $\bm{\iota}_{n-1}^{\alpha}$ and $\bm{\iota}_{n}^{\beta}$. Higher $\rho$ between $\bm{\iota}_{n-1}^{\alpha}$ and $\bm{\iota}_{n}^{\beta}$ means that $\bm{\iota}_{n}^{\beta}$ is more possibility of being consistent with $\bm{\iota}_{n-1}^{\alpha}$ in time sequence, and that the learned result from $\beta$ is eligible to preserve $\alpha$'s learning procedures. With the defined \emph{similarity}, we expect \eg the vehicle under city-street-driving can only learn from vehicles under similar driving patterns rather than to learn from highway-driving vehicles.

We introduce a pre-defined coefficient $\mu$ as a threshold for $\rho$. When the vehicle $\alpha$ is detected to be with anomaly at the $n$-th instant, the \emph{similarity} $\rho$ between $\alpha$'s most recent available learned result $\bm{\iota}_{n-1}^{\alpha}$ and all possible learned results from anomaly-free vehicles in the same network will be calculated. All the eligible learned results satisfying $\rho\ge\mu$ will be averaged as $\bm{\iota}_{n}^{*}$, and the learning iteration for $\alpha$ at the $n$-th instant will be preserved and continued as:

\begin{equation}\label{equation:model parameter renew for Alpha}
\begin{split}
\bm{\omega}_{n}^{\alpha}=\bm{\omega}_{n-1}^{\alpha}+\bm{\iota}_{n}^{*}\\
\end{split}
\end{equation}

\begin{figure}[!htbp]
  \centering
  \includegraphics[width=0.45\textwidth]{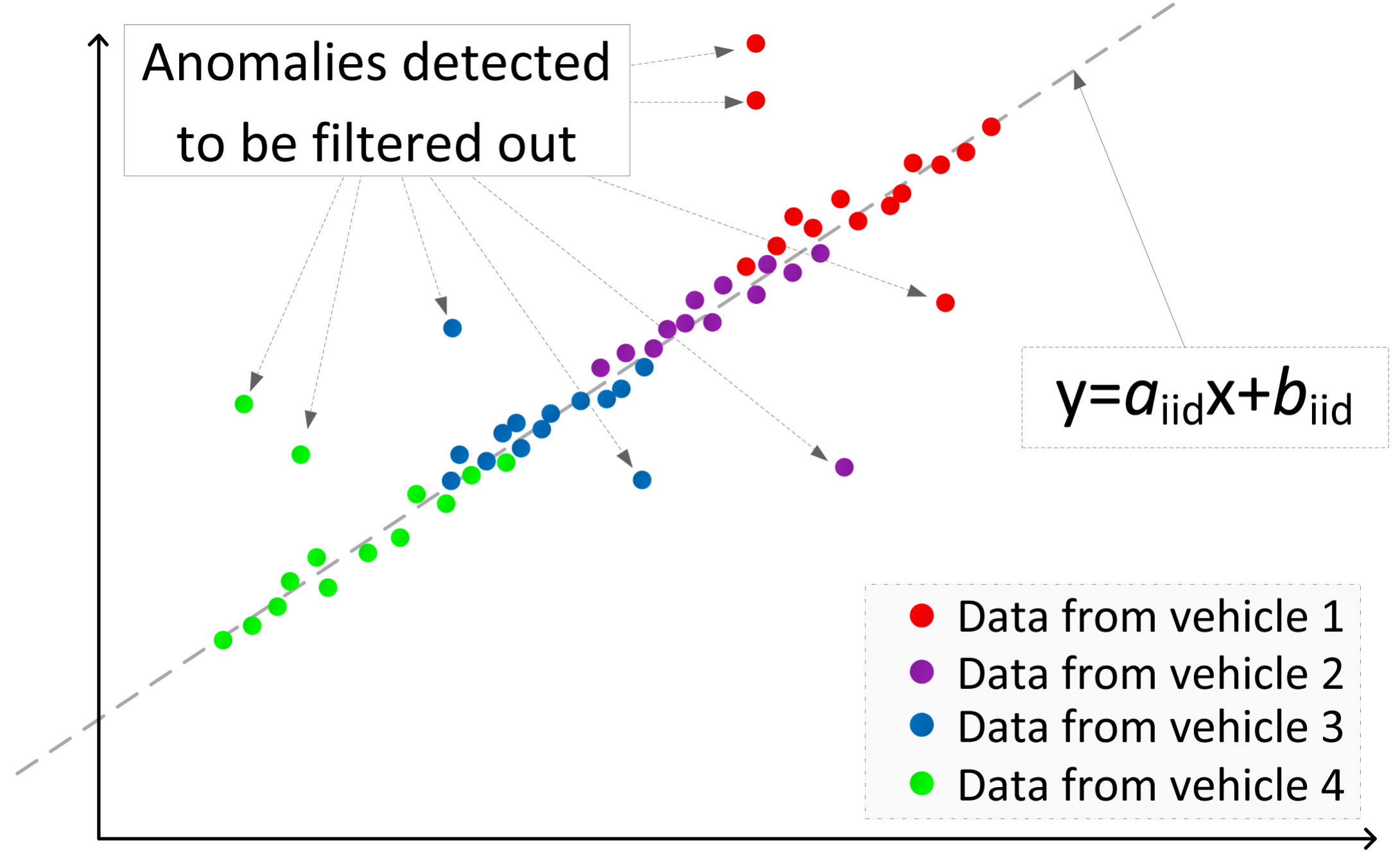}\\
  \caption{\label{fig: IID_dataset} Modeling for i.i.d data distribution}
\end{figure}

It should be noticed that, though we propose to address the challenge of heterogeneous data distribution, our sharing policy is also applicable to i.i.d datasets, where one single estimation model can cover all the vehicles. When applying the sharing policy on i.i.d data shown in Figure~\ref{fig: IID_dataset}, all the vehicles, including those suffering anomalies, will learn on data that contribute to the same model, and eventually every vehicle will be assigned with the same model.

\section{Implementation and experimental results}\label{section: experiments}

In this section, six FL approaches, including the proposed one, are implemented under the circumstance of heterogeneous data distribution, and performance comparisons show our approach gains significant advantages in model estimation accuracy, without transmitting raw data that induces direct privacy vulnerability nor increasing the time complexity in algorithm execution. The implementation details are shown in the following.

\subsection{Evaluation environment and experiment settings}

\subsubsection{Data and methods used in experiments}

The implementation is conducted in MATLAB R2018b based on the Windows 10 operating system, using the open-source database of \textit{Battery and heating data in real driving cycles}~\cite{DataSource} for performance verification. 26 vehicle trip datasets are employed, where 4000 consecutive records are used as the training dataset, and the following 2000 records are used as the testing dataset for each trip. Details of data features are shown in Table~\ref{table: input-output features}.

\begin{figure}[!htbp]
  \centering
  \includegraphics[width=0.7\textwidth]{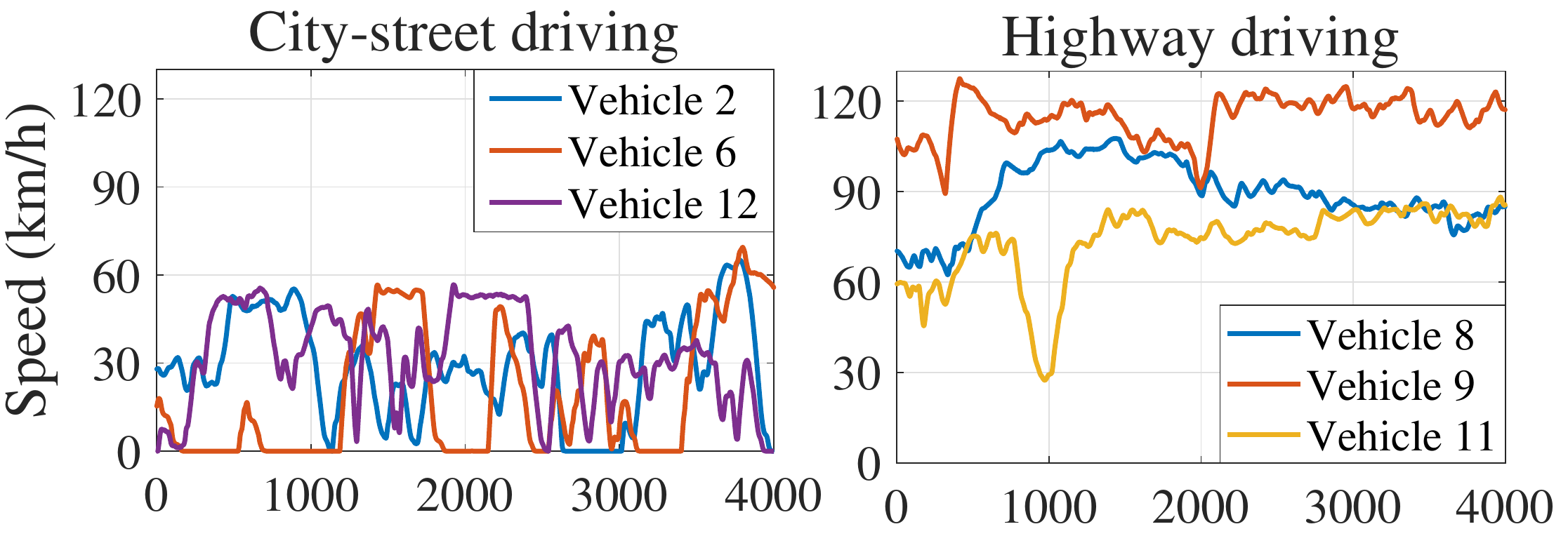}\\
  \caption{\label{fig: different driving scenarios}Examples of vehicles under different driving patterns}
\end{figure}

The 26 vehicle trips employed are under two different driving patterns: city street driving and highway driving, distinguished by their speed profile, as shown in Figure~\ref{fig: different driving scenarios}. We intentionally include such distinctive driving patterns to increase heterogeneity of data distribution, and adopt a considerable number of input features to gain heterogeneity further.

\begin{table}[htp]
\centering
\caption{\label{table: input-output features}Input and output features description}
\resizebox{0.7\textwidth}{!}{
\begin{tabular}{|l|l|l|}
\hline
data & \multicolumn{2}{c|}{features}\\
\hline
\multirow{2}{*}{input} & \multicolumn{2}{l|}{\tabincell{l}{constant; trip length (m); velocity (km/h); elevation (m); throttle (\%);\\ motor torque (Nm); longitudinal acceleration (m/s$^{2}$); regenerative\\ braking signal (0 or 1); battery voltage (V); battery current (A);\\ battery temperature ($^{\circ}$C); max battery temperature ($^{\circ}$C); min SoC (\%);\\ max SoC (\%); heating power CAN (kW); heating power LIN (W);\\ requested heating power (W); air condition power (kW); heater signal\\ (0 or 1); heater voltage (V); heater current (A)}} \\
\cline{2-3}
& \tabincell{c}{temperature\\($^{\circ}$C)} & \tabincell{l}{ambient temperature; ambient sensor; coolant heater\\ core; requested coolant; coolant inlet; heat exchanger;\\ cabin sensor} \\
\hline
output & \multicolumn{2}{c|}{remaining EV battery volume (\%)}\\
\hline
\end{tabular}
}
\end{table}

We implement six approaches in the experiments, including the proposed extended federated-learning (E-FL), unconditional sharing federated-learning method (U-FL), federated-learning with assumption on membership relationship~\cite{FLChallenge_MTL} (AS-FL), Federated-learning with restricted learned results~\cite{FLChallenge_LimitedLocalImpact} (R-FL), and also with federated-learning only equipped with the developed anomaly detection (AD-FL) and federated-learning equipped only with the devised sharing policy (S-FL). The last two methods are implemented to show how the developed sharing policy and anomaly detection perform separately.

\subsubsection{Model structure}

\begin{figure}[!htbp]
  \centering
  \includegraphics[width=0.6\textwidth]{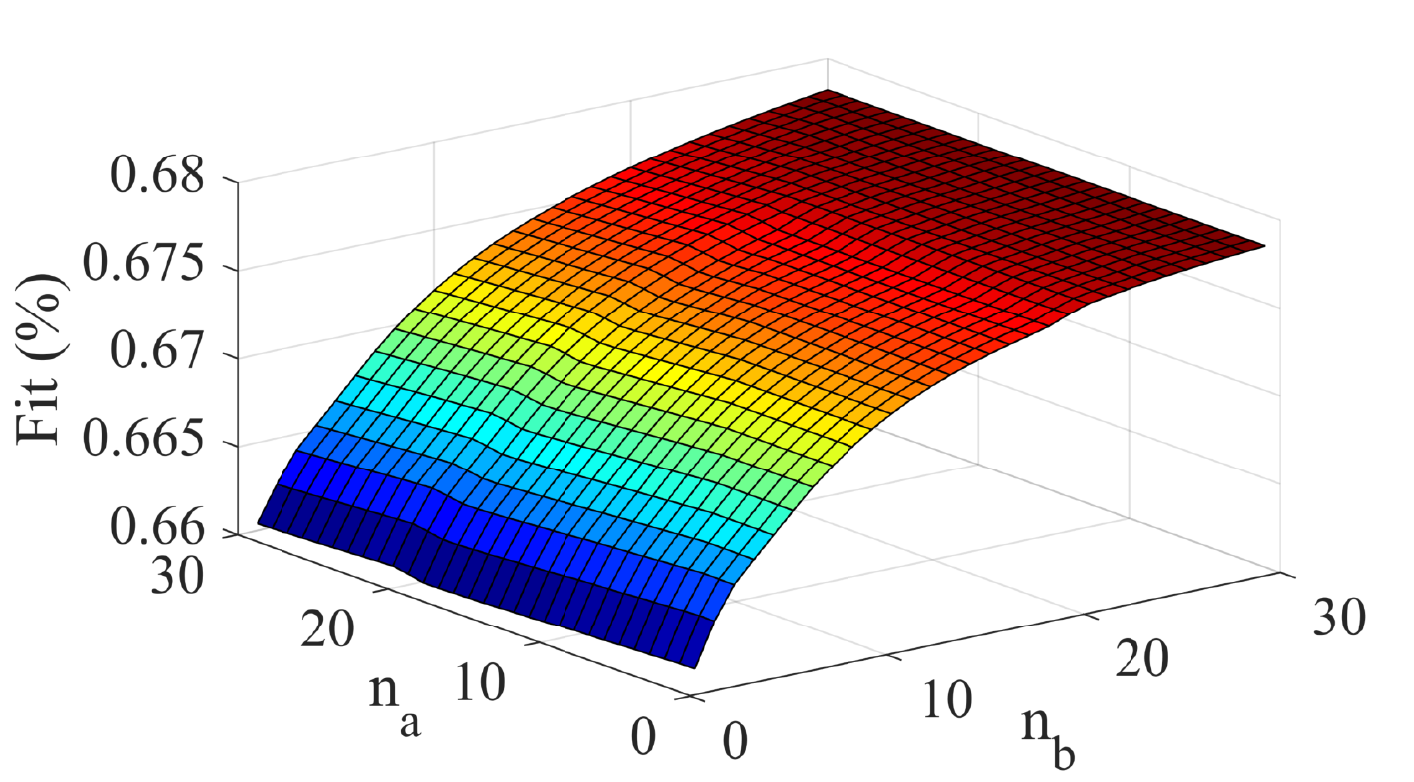}\\
  \caption{\label{fig: order} Model performance with different orders}
\end{figure}

All the six of implemented approaches follow an adaptive Auto-Regressive eXogenous (ARX) model~\cite{ARXmodel} shown in equation~\ref{equation: ARX model}, where the adaptive features~\cite{AdaptiveLearningFeature2} are leveraged to improve the prediction accuracy. The ARX model's order used in experiments is selected from 1 to 30. The selection is conducted via evaluating model performance under different orders as shown in Figure~\ref{fig: order}, which is measured by the \textit{fit} index~\cite{ModelFitIndex} as: $fit(.)=(1-\frac{||\hat y-y||}{||y-\overline y||})\cdot 100\%$, where $y$, $\overline y$, and $\hat y$ are the system out, its mean value, and the output estimated by the model, respectively. This order selection is achieved based on an extra training dataset with 5000 records.

\begin{equation}\label{equation: ARX model}
\begin{split}
A(q)y(t)=\emph{\textbf{B}}(q)\textbf{x}(t)+e(t)\\
\textbf{x}(t)=(x_{1}(t),x_{2}(t),...,x_{n}(t))^{T}\\
A(q)=1+\sum\limits_{k=1}^{n_{a}}a_{k}q^{-k}\\
\emph{\textbf{B}}(q)=(B(q)_{1},B(q)_{2},...,B(q)_{n})\\
B(q)_{i}=1+\sum\limits_{k=1}^{n_{b}}b^{i}_{k}q^{-k}, i\in\{1,2,...,n\}\\
\end{split}
\end{equation}
where $y(t)$ and $\textbf{x}(t)$ represent the system output and input respectively at the $k$-th instant, $n_{a}$ denotes the order of system output, $n_{b}$ is the order of system input, $n$ is the number of input attributes, $e$ is the white noise and $q^{-1}$ is the delay operator.

\begin{figure*}[!htbp]
  \centering
  \includegraphics[width=1\textwidth]{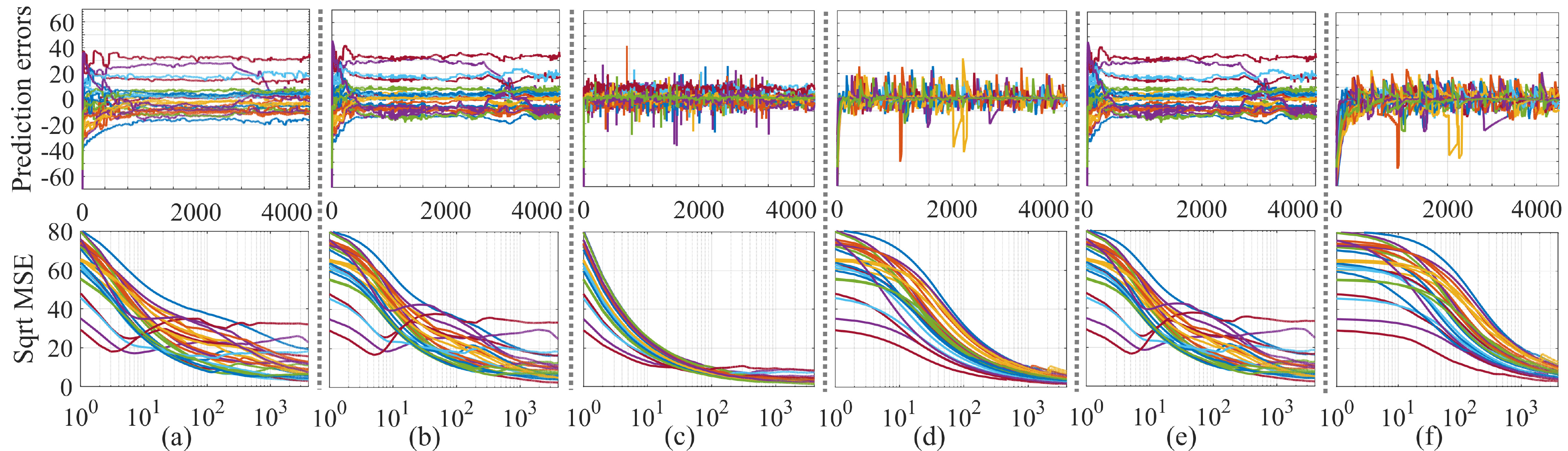}\\
  \caption{\label{fig: integrated results}Prediction performance and convergence of sqrt MSE on prediction for all considered approaches}
\end{figure*}

High model order brings improved model as shown in Figure~\ref{fig: order}, yet the highest-order model in the considered range merely brings a 1.73\% increased \emph{fit} compared with the lowest-order model. Higher model order does not lead to significant improvement since there is already a considerable number of input attributes in the dataset. To reduce the computation burden, we use a 1-order model in the following experiments.

\subsubsection{Evaluation criterion}

With the model structure established, experiments can be conducted on the training dataset to find the optimal model parameters for each considered method, and then the obtained parameters are leveraged on the testing dataset to estimate energy consumption. In the comparison of different methods, we employ the same criterion of sqrt mean squared error (MSE) to evaluate the estimation accuracy of the learned models. The convergence performance of each approach is also displayed in the training phase measured by the learning iterations needed to converge.

\subsection{Experimental results: prediction and convergence on the training dataset}

This section presents the prediction and convergence performance of different methods on the training dataset. Figure~\ref{fig: integrated results}-(a) shows how it works when the sharing of learned results is under no constraints and one single model is generated to cover all vehicles. In this figure, each curve represents one vehicle's performance, and the upper and lower part show the prediction errors and sqrt MSE on prediction at each iteration, respectively. Figure~\ref{fig: integrated results}-(a) show that none of the vehicles obtain desirable prediction performance, with each of their prediction error curves fluctuating significantly over the training procedures. Such results indicate that the single model generated fails in the presence of data distribution heterogeneity.

The prediction performance under R-FL is shown in Figure~\ref{fig: integrated results}-(b), where smaller prediction errors can be observed compared with U-FL shown in Figure~\ref{fig: integrated results}-(a). The sqrt MSE on prediction converges faster compared with U-FL. Nevertheless, such results are far from satisfying since none of the vehicles' prediction error curves could stably stay around zero. Compared with U-FL, the influence of data heterogeneity on modeling is weakened rather than efficiently reduced.

AS-FL provides much better predictions than U-FL and R-FL, as observed in Figure~\ref{fig: integrated results}-(c). For each of the 26 vehicles, prediction errors are concentrated around zeros with fewer deviations than U-FL and R-FL, and the sqrt MSE on prediction converges better than U-FL or R-FL. Those results imply that data heterogeneity is solved better, and a more efficient estimation model is generated for all the vehicles. However, this model efficiency pays the price of privacy: raw vehicle data is transmitted to the learning engine directly in this approach, inducing vulnerability of privacy-sensitive information disclosing, \eg driver location and address.

\begin{figure}[!htbp]
  \centering
  \includegraphics[width=1\textwidth]{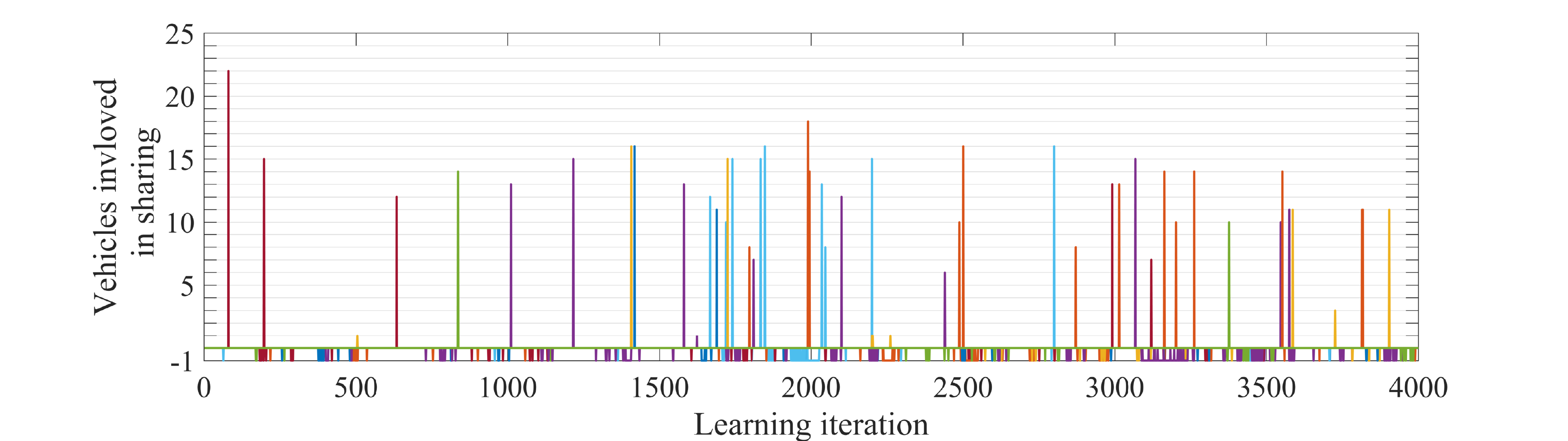}\\
  \caption{\label{fig: sharing results}Anomaly detection and sharing results in extended FL}
\end{figure}

Evaluation results of the developed E-FL are shown in Figure~\ref{fig: integrated results}-(d), where the prediction error curves concentrate around 0, with fluctuations of limited magnitude (about 20). Note that this prediction performance is obtained without the transmit of raw data compared with AS-FL. Those results infer that heterogeneous data distribution is handled with higher efficiency compared with U-FL and R-FL. For the 4000$\times$26 entries in the training dataset, 49 of them are detected as anomalies and filtered out in learning procedures. Amongst the anomaly events detected, 45 of them continue their learning procedures via extracting learned results shared from other vehicles. The anomaly detection protects the learning procedures from obstacles in convergence, while the sharing policy offers a chance for vehicles under anomaly to continue their learning procedures. The anomaly detection and sharing results in E-FL are shown in Figure~\ref{fig: sharing results} (with $\rho=0.2$, $K=10$, $\theta=3$), where different vehicles are assigned with different colored curves, and the vertical axis represents vehicles involved in sharing when an individual vehicle is detected to be with anomaly. 0 on the vertical axis means anomaly-free and model renew only with one's local data, and -1 denotes that no renew is needed since the prediction error is below a predefined threshold.

\begin{table*}[htp]
\centering
\caption{\label{table: overall results}Summarized performance results of six considered methods on training datasets}
\resizebox{1\textwidth}{!}{
\begin{tabular}{ccccccccccccccccccccccccccccc}
\toprule
\multirow{2}{*}{method} & \multirow{2}{*}{criteria} & \multicolumn{26}{c}{trip number} & \multirow{2}{*}{\tabincell{c}{averaged\\result}}\\
\cmidrule{3-28}
& & 1 & 2 & 3 & 4 & 5 & 6 & 7 & 8 & 9 & 10 & 11 & 12 & 13 & 14 & 15 & 16 & 17 & 18 & 19 & 20 & 21 & 22 & 23 & 24 & 25 & 26 &   \\
\toprule
\multirow{2}{*}{E-FL} & sqrt $MSE$ & 6.36 & 4.55 & 4.57 & 4.70 & 4.28 & 3.53 & 3.71 & 2.84 & 6.34 & 8.20 & 3.97 & 3.14 & 2.78 & 2.17 & 3.36 & 4.41 & 4.55 & 5.81 & 4.04 & 2.50 & 1.82 & 3.44 & 5.13 & 7.16 & 2.05 & 4.02 &  4.21 \\
\cmidrule{2-29}
& $I_{C}$ & 150 & 150 & 50 & 150 & 400 & 220 & 150 & 200 & 70 & 200 & 100 & 100 & 60 & 80 & 50 & 100 & 450 & 300 & 100 & 180 & 140 & 450 & 200 & 250 & 30 & 100 & 170.38  \\
\midrule
\multirow{2}{*}{U-FL} & sqrt $MSE$ & 19.18 & 12.79 & 3.72 & 22.66 & 8.25 & 3.38 & 2.80 & 6.39 & 8.01 & 6.15 & 12.12 & 10.99 & 3.81 & 15.56 & 5.67 & 9.07 & 6.67 & 11.54 & 6.46 & 18.07 & 32.01 & 4.39 & 12.83 & 7.74 & 7.40 & 6.28 & 10.15  \\
\cmidrule{2-29}
& $I_{C}$ & 500 & 100 & 200 & 550 & 350 & 300 & 500 & 250 & 400 & 750 & 50 & 400 & 50 & 40 & 160 & 400 & 600 & 700 & 650 & 150 & 650 & 600 & 1000 & 900 & 600 & 900 & 451.92  \\
\midrule
\multirow{2}{*}{AS-FL} & sqrt $MSE$ & 3.01 & 4.45 & 3.00 & 7.50 & 3.13 & 2.83 & 2.14 & 2.90 & 4.41 & 2.14 & 5.78 & 3.43 & 2.80 & 4.53 & 3.12 & 3.61 & 2.40 & 2.61 & 2.68 & 7.45 & 8.47 & 2.56 & 3.11 & 2.33 & 2.75 & 1.48 &  3.64 \\
\cmidrule{2-29}
& $I_{C}$ & 5 & 10 & 10  & 20 & 4  & 20  & 60  & 5  & 25  & 50  & 5  & 120  & 80  & 100  & 3 & 5 & 10  & 5 & 10  & 200  & 20  & 5 & 100  & 10  & 10  & 10  &  34.69  \\
\midrule
\multirow{2}{*}{R-FL} & sqrt $MSE$ & 16.13 & 10.62  & 4.85  & 24.84  & 6.88  & 4.15  & 2.50  & 6.93  & 4.78  & 4.22  & 11.66  & 12.73  & 4.55  & 16.56  & 6.08  & 9.21  & 4.71  & 9.59  & 7.83  & 18.55  & 33.35  & 4.24  & 10.63  & 5.10  & 10.39  & 7.34  &  9.94  \\
\cmidrule{2-29}
& $I_{C}$ & 400  & 250  & 250  & 400  & 350  & 400  & 350  & 400  & 450  & 400  & 600  & 500  & 500  & 400  & 500  & 400  & 400  & 300  & 350  & 500  & 450  & 200  & 400  & 500  & 600  & 450  & 411.53   \\
\midrule
\multirow{2}{*}{AD-FL} & sqrt $MSE$ & 16.12  & 10.61  & 4.86  & 24.85  & 6.86  & 4.17  & 2.51  & 6.94  & 4.76  & 4.22  & 11.65  & 12.71  & 4.56  & 16.57  & 6.07  & 9.19  & 4.70  & 9.57  & 7.84  & 18.57  & 33.36  & 4.27  & 10.62  & 5.10  & 10.35  & 7.36  &  9.94  \\
\cmidrule{2-29}
& $I_{C}$ & 400  & 250  & 250  & 300  & 350  & 300  & 350  & 400  & 450  & 400  & 20  & 500  & 350  & 350  & 500  & 100  & 400  & 300  & 400  & 300  & 400  & 200  & 450  & 600  & 600  & 450  &  360.38  \\
\midrule
\multirow{2}{*}{S-FL} & sqrt $MSE$ & 11.41  & 8.82  & 7.80  & 8.39  & 8.26  & 6.38  & 6.98  & 4.79  & 10.36  & 13.10  & 7.31  & 6.37  & 5.14  & 4.06  & 6.51  & 7.96  & 9.26  & 10.84  & 6.99  & 4.75  & 3.00  & 4.93  & 9.97  & 11.64  & 4.10  & 7.75  & 7.57   \\
\cmidrule{2-29}
& $I_{C}$ & 400  & 350  & 400  & 700  & 350  & 250  & 200  & 250  & 1200  & 750  & 150  & 150  & 150  & 150  & 200  & 500  & 400  & 300  & 200  & 200  & 750  & 500  & 700  & 700  & 200  & 600  &  411.53  \\
\toprule
\end{tabular}
}
\end{table*}

Prediction performance under AD-FL is shown in Figure~\ref{fig: integrated results}-(e). Though AD-FL's learning procedures are protected against anomalies, the prediction errors can not concentrate around zero, and the sqrt MSE of prediction for several vehicles does not converge well. For most of the vehicle prediction error curves, significant fluctuations exist during all the training iterations. Such results suffer from the same problem as in U-FL and R-FL, where the use of one single model can not cover all the considered vehicles, especially when heterogeneity exists amongst different vehicles.

\begin{table}[htp]
\centering
\caption{\label{table: complexity}Time complexity and raw data transmitting for different approaches}
\resizebox{0.5\textwidth}{!}{
\begin{tabular}{|c|c|c|}
\hline
approach & \tabincell{c}{time complexity\\to the amount of\\network participants} & \tabincell{l}{sharing raw\\data or not}\\
\hline
unconditional sharing FL & O(n$^{2}$) & \xmark \\
\hline
FL with restricted learned result & O(n$^{2}$) & \xmark \\
\hline
FL with assumed relationship & O(n) & \cmark \\
\hline
extended FL & O(n) to O(n$^{2}$) & \xmark \\
\hline
FL with anomaly detection only & O(n$^{2}$) & \xmark \\
\hline
FL with sharing policy only & O(n) to O(n$^{2}$) & \xmark \\
\hline
\end{tabular}
}
\end{table}

Figure~\ref{fig: integrated results}-(f) shows prediction results under S-FL, where the prediction errors are more concentrated around zero than those in Figure~\ref{fig: integrated results}-(e), indicating a more efficient sharing of results amongst vehicles that enhance their learning procedures. However, compared with E-FL, this execution bears more fluctuations in prediction iterations, implying that anomalies are free from constraints and bring disturbances in learning procedures.

\begin{table*}[htp]
\centering \caption{\label{table: Testing results}Experimental results on testing datasets measured by sqrt MSE}
\resizebox{0.98\textwidth}{!}{
\begin{tabular}{cccccccccccccccccccccccccccc}
\toprule
\multirow{2}{*}{method} & \multicolumn{26}{c}{trip number} & \multirow{2}{*}{\tabincell{c}{averaged\\result}} \\
\cmidrule{2-27}
 & 1 & 2 & 3 & 4 & 5 & 6 & 7 & 8 & 9 & 10 & 11 & 12 & 13 & 14 & 15 & 16 & 17 & 18 & 19 & 20 & 21 & 22 & 23 & 24 & 25 & 26 &   \\
\toprule
E-FL & 9.35 & 7.55 & 4.48 & 7.34 & 5.06 & 3.83 & 5.46 & 3.90 & 4.94 & 10.75 & 5.64 & 4.61 & 3.63 & 3.27 & 5.27 & 5.50 & 6.12 & 5.96 & 6.23 & 2.11 & 2.58 & 3.52 & 6.41 & 5.79 & 2.88 & 5.13 & 5.28\\
\midrule
U-FL & 20.97 & 14.58 & 3.58 & 22.58 & 8.28 & 10.38 & 2.76 & 2.78 & 11.22 & 11.25 & 15.76 & 12.69 & 2.61 & 14.35 & 6.69 & 13.08 & 7.23 & 10.98 & 8.60 & 25.00 & 30.60 & 6.08 & 10.81 & 4.98 & 8.72 & 6.57 & 11.28\\
\midrule
AS-FL & 4.63 & 4.58 & 4.35 & 8.24 & 2.61 & 3.30 & 2.65 & 2.74 & 5.11 & 3.25 & 4.80 & 4.87 & 2.69 & 3.94 & 4.55 & 2.91 & 2.44 & 3.04 & 3.12 & 8.19 & 11.04 & 2.66 & 3.29 & 2.50 & 3.21 & 1.73 & 4.09\\
\midrule
R-FL & 18.61 & 12.51 & 4.97 & 24.97 & 6.32 & 12.03 & 2.66 & 3.53 & 8.90 & 9.23 & 14.14 & 13.58 & 4.44 & 15.43 & 5.40 & 10.00 & 5.18 & 8.08 & 9.57 & 27.14 & 32.27 & 9.24 & 8.21 & 3.20 & 11.65 & 7.45 & 11.10\\
\midrule
AD-FL & 19.64 & 13.05 & 5.42 & 24.37 & 6.50 & 14.23 & 2.92 & 3.60 & 8.91 & 10.98 & 14.03 & 13.30 & 4.67 & 15.52 & 5.42 & 10.72 & 5.82 & 8.70 & 9.48 & 29.33 & 32.14 & 9.89 & 8.60 & 3.50 & 12.26 & 7.35 & 11.55\\
\midrule
S-FL & 14.09 & 11.29 & 7.15 & 10.46 & 7.25 & 5.57 & 8.18 & 4.38 & 7.74 & 15.72 & 8.48 & 7.18 & 3.72 & 5.43 & 8.03 & 8.36 & 9.61 & 7.56 & 9.40 & 2.80 & 4.08 & 5.37 & 9.19 & 8.93 & 3.63 & 8.03 & 7.75\\
\bottomrule
\end{tabular}
}
\end{table*}

Table~\ref{table: overall results} presents the numerical comparison of results from different approaches, including the results of sqrt MSE on prediction for the whole training iterations and iterations needed to converge (denoted by $I_{C}$). The time complexity for the execution of each approach is listed in Table~\ref{table: complexity}. It can be summarized from Table~\ref{table: overall results} that AS-FL provides the least prediction error with the least iterations to reach convergence. Nevertheless, the sharing of raw data in AS-FL exposes privacy vulnerability and makes it not the optimal choice for customers. In contrast, our approach could offer significant promotion in prediction and convergence performance compared with U-FL, R-FL, AD-FL, and S-FL, without transmitting raw data or increasing time complexity. When we average the 26-vehicle experiment's sqrt MSE and convergence iteration in Table~\ref{table: overall results}, our approach shows 58.5\%, 57.6\%, 57.6\%, 44.3\% reduced sqrt MSE, and 62.3\%, 58.6\%, 52.7\%, 58.6\% reduced convergence iterations compared with U-FL, R-FL, AD-FL, S-FL, respectively.

\subsection{Experimental results: estimation performance on the testing dataset}

Given each method's trained model parameters, all the considered methods are tested using the testing dataset for energy consumption estimation. It should be noted that long-term energy estimation requires prospective information~\cite{eRout9}, \eg vehicle speed profiles in the future. In this experiment, we treat data attributes \eg vehicle speed, acceleration signal as profiles that the driver will following during the upcoming trip, and provides energy consumption estimation based on those profiles.

The testing results are shown in Table~\ref{table: Testing results} using the evaluation criterion of sqrt MSE on prediction error. The averaged results show that, our approach reduces prediction error by 53.19\%, 52.43\%, 54.29\%, and 31.87\% compared with U-FL, R-FL, AD-FL, and S-FL, respectively. Though AS-FL shows more prediction error reduced, this reduction is not significant and is gained at the price of privacy, where raw vehicle data is directly transmitted and the vulnerability of privacy-sensitive information leakage is induced.

\section{Conclusions}\label{section: conclustion}

This paper proposes a federated-learning-based approach contributing to energy-efficient route planning for EV networks. Our approach leverages shared information from vehicles to enhance network learning while avoids transmitting raw data during learning procedures. More importantly, the new scheme extends the federated-learning structure to gain learning robustness against data distribution heterogeneity in EV networks. Experimental results verify the new approach's efficiency, which shows high energy estimation accuracy is obtained confronted with heterogeneous vehicle-data distributions. Beyond the scenario of EV networks, the proposed approach also provides a possible solution for various network learning applications where each participant with individual features collaborates to extract knowledge from data.

\section*{Acknowledgement}

This work was supported by the project `Privacy-Protected Machine Learning for Transport Systems' of Area of Advance Transport and Chalmers AI Research Centre (CHAIR). The computations were enabled by resources provided by the Swedish National Infrastructure for Computing (SNIC) at C3SE partially funded by the Swedish Research Council through grant agreement no. 2018-05973. I also thank Elad Schiller for important discussions and helping to improve the presentation.

\bibliographystyle{IEEEtran}
\bibliography{bibfile}

\begin{thebibliography}{10}
\providecommand{\url}[1]{#1}
\csname url@samestyle\endcsname
\providecommand{\newblock}{\relax}
\providecommand{\bibinfo}[2]{#2}
\providecommand{\BIBentrySTDinterwordspacing}{\spaceskip=0pt\relax}
\providecommand{\BIBentryALTinterwordstretchfactor}{4}
\providecommand{\BIBentryALTinterwordspacing}{\spaceskip=\fontdimen2\font plus
\BIBentryALTinterwordstretchfactor\fontdimen3\font minus
  \fontdimen4\font\relax}
\providecommand{\BIBforeignlanguage}[2]{{%
\expandafter\ifx\csname l@#1\endcsname\relax
\typeout{** WARNING: IEEEtran.bst: No hyphenation pattern has been}%
\typeout{** loaded for the language `#1'. Using the pattern for}%
\typeout{** the default language instead.}%
\else
\language=\csname l@#1\endcsname
\fi
#2}}
\providecommand{\BIBdecl}{\relax}
\BIBdecl

\bibitem{AcceptanceSurvey1}
E.~A. {Prasetio}, P.~{Fajarindra Belgiawan}, L.~T. {Anggarini},
  D.~{Novizayanti}, and S.~{Nurfatiasari}, ``Acceptance of electric vehicle in
  indonesia: Case study in bandung,'' in \emph{6th International Conference on
  Electric Vehicular Technology}, Bali, Indonesia, 2019, pp. 63--71.

\bibitem{AcceptanceSurvey2}
J.~M. M{\"u}ller, ``Comparing technology acceptance for autonomous vehicles,
  battery electric vehicles, and car sharing—a study across europe, china,
  and north america,'' \emph{Sustainability}, vol.~11, no.~16, pp. 1--17, 2019.

\bibitem{RangeAnxiety3}
A.~T. {Thorgeirsson}, S.~{Scheubner}, S.~{Fünfgeld}, and F.~{Gauterin}, ``An
  investigation into key influence factors for the everyday usability of
  electric vehicles,'' \emph{IEEE Open Journal of Vehicular Technology},
  vol.~1, pp. 348--361, 2020.

\bibitem{HurdlesForEV}
G.~{Tamai}, ``What are the hurdles to full vehicle electrification?''
  \emph{IEEE Electrification Magazine}, vol.~7, no.~1, pp. 5--11, 2019.

\bibitem{e-route-recently}
J.~Wang, A.~Elbery, and H.~A. Rakha, ``A real-time vehicle-specific eco-routing
  model for on-board navigation applications capturing transient vehicle
  behavior,'' \emph{Transportation Research Part C: Emerging Technologies},
  vol. 104, pp. 1--21, 2019.

\bibitem{eRout1}
K.~{Kraschl-Hirschmann} and M.~{Fellendorf}, ``Estimating energy consumption
  for routing algorithms,'' in \emph{IEEE Intelligent Vehicles Symposium},
  Alcala de Henares, Spain, 2012, pp. 258--263.

\bibitem{eRout2}
G.~A. W.~D. Baum~Moritz, Dibbelt~Julian, ``Towards route planning algorithms
  for electric vehicles with realistic constraints,'' \emph{Computer Science -
  Research and Development}, vol.~31, pp. 105--109, 2016.

\bibitem{eRout3}
S.~{Grubwinkler}, M.~{Hirschvogel}, and M.~{Lienkamp}, ``Driver- and
  situation-specific impact factors for the energy prediction of evs based on
  crowd-sourced speed profiles,'' in \emph{IEEE Intelligent Vehicles Symposium
  Proceedings}, Dearborn, MI, USA, 2014, pp. 1069--1076.

\bibitem{eRout4}
M.~{Neaimeh}, G.~A. {Hill}, Y.~{Hübner}, and P.~T. {Blythe}, ``Routing systems
  to extend the driving range of electric vehicles,'' \emph{IET Intelligent
  Transport Systems}, vol.~7, no.~3, pp. 327--336, 2013.

\bibitem{eRout5}
A.~{Liebscher}, M.~{Krumnow}, J.~{Krimmling}, F.~{Hanisch}, and B.~{Baker},
  ``Energy-efficient routing strategies based on real-time data of a local
  traffic management center,'' in \emph{International Conference on Models and
  Technologies for Intelligent Transportation Systems}, Budapest, Hungary,
  2015, pp. 74--80.

\bibitem{eRout6}
A.~E. T.~M. Masikos~Michail, Demestichas~Konstantinos, ``Mesoscopic forecasting
  of vehicular consumption using neural networks,'' \emph{Soft Computing},
  vol.~19, pp. 145--156, 2013.

\bibitem{eRout7}
M.~V. D.~F. Ferreira Joao~C and J.~L. Afonso, ``Data mining approach for range
  prediction of electric vehicle,'' in \emph{Conference on Future Automotive
  Technology-Focus Electromobility}, Munich, Germany, 2012, pp. 1--15.

\bibitem{eRout8}
A.~M. {Bozorgi}, M.~{Farasat}, and A.~{Mahmoud}, ``A time and energy efficient
  routing algorithm for electric vehicles based on historical driving data,''
  \emph{IEEE Transactions on Intelligent Vehicles}, vol.~2, no.~4, pp.
  308--320, 2017.

\bibitem{eRout13}
R.~{Yaqub} and Y.~{Cao}, ``Smartphone-based accurate range and energy efficient
  route selection for electric vehicle,'' in \emph{IEEE International Electric
  Vehicle Conference}, Greenville, SC, USA, 2012, pp. 1--5.

\bibitem{DrivingAndStationary}
C.~{Schuss}, T.~{Fabritius}, B.~{Eichberger}, and T.~{Rahkonen},
  ``Energy-efficient routing of electric vehicles with integrated photovoltaic
  installations,'' in \emph{IEEE International Instrumentation and Measurement
  Technology Conference}, Dubrovnik, Croatia, 2020, pp. 1--6.

\bibitem{InputFactorsMore}
Z.~{Yi} and P.~H. {Bauer}, ``Adaptive multiresolution energy consumption
  prediction for electric vehicles,'' \emph{IEEE Transactions on Vehicular
  Technology}, vol.~66, no.~11, pp. 10\,515--10\,525, 2017.

\bibitem{BellmanFordBasedMethod}
C.~{De Cauwer}, W.~{Verbeke}, J.~{Van Mierlo}, and T.~{Coosemans}, ``A model
  for range estimation and energy-efficient routing of electric vehicles in
  real-world conditions,'' \emph{IEEE Transactions on Intelligent
  Transportation Systems}, vol.~21, no.~7, pp. 2787--2800, 2020.

\bibitem{FederatedLearningOnIIDdata}
Y.~Zhao, M.~Li, L.~Lai, N.~Suda, D.~Civin, and V.~Chandra, ``Federated learning
  with non-iid data,'' \emph{CoRR}, vol. abs/1806.00582, 2018.

\bibitem{FederatedLearning}
S.~A. {Rahman}, H.~{Tout}, H.~{Ould-Slimane}, A.~{Mourad}, C.~{Talhi}, and
  M.~{Guizani}, ``A survey on federated learning: The journey from centralized
  to distributed on-site learning and beyond,'' \emph{IEEE Internet of Things
  Journal}, vol.~0, no.~0, pp. 1--23, 2020.

\bibitem{Federated-learningAndPrivacyProtection}
Q.~Li, Z.~Wen, and B.~He, ``Federated learning systems: Vision, hype and
  reality for data privacy and protection,'' \emph{CoRR}, vol. abs/1907.09693,
  2019.

\bibitem{FLChallenge_nonIID2}
E.~Jeong, S.~Oh, H.~Kim, J.~Park, M.~Bennis, and S.-L. Kim,
  ``Communication-efficient on-device machine learning: Federated distillation
  and augmentation under non-iid private data,'' \emph{CoRR}, vol.
  abs/1811.11479, 2018.

\bibitem{FLChallenge_MTL}
V.~Smith, C.-K. Chiang, M.~Sanjabi, and A.~S. Talwalkar, ``Federated multi-task
  learning,'' in \emph{Advances in Neural Information Processing Systems},
  I.~Guyon, U.~V. Luxburg, S.~Bengio, H.~Wallach, R.~Fergus, S.~Vishwanathan,
  and R.~Garnett, Eds., vol.~30, 2017, pp. 4424--4434.

\bibitem{FLChallengs_NNModel}
H.~Wu, C.~Chen, and L.~Wang, ``A theoretical perspective on differentially
  private federated multi-task learning,'' \emph{CoRR}, vol. abs/2011.07179,
  2020.

\bibitem{FLChallenge_block}
H.~Eichner, T.~Koren, B.~McMahan, N.~Srebro, and K.~Talwar, ``Semi-cyclic
  stochastic gradient descent,'' in \emph{Proceedings of the 36th International
  Conference on Machine Learning, 9-15 June, Long Beach, California, {USA}},
  vol.~97, 2019, pp. 1764--1773.

\bibitem{FLChallenge_Clustering}
F.~{Sattler}, K.~R. {Müller}, and W.~{Samek}, ``Clustered federated learning:
  Model-agnostic distributed multitask optimization under privacy
  constraints,'' \emph{IEEE Transactions on Neural Networks and Learning
  Systems}, vol. early access, pp. 1--13, 2020.

\bibitem{FlChallenge_Clustering2}
A.~Ghosh, J.~Hong, D.~Yin, and K.~Ramchandran, ``Robust federated learning in a
  heterogeneous environment,'' \emph{CoRR}, vol. abs/1906.06629, 2019.

\bibitem{FLChallenge_LimitedLocalImpact}
A.~K. Sahu, T.~Li, M.~Sanjabi, M.~Zaheer, A.~Talwalkar, and V.~Smith, ``On the
  convergence of federated optimization in heterogeneous networks,''
  \emph{arXiv preprint arXiv:1812.06127}, vol.~3, 2018.

\bibitem{FLChallenge_LimitLearnedResultImpact2}
L.~Li, W.~Xu, T.~Chen, G.~B. Giannakis, and Q.~Ling, ``Rsa: Byzantine-robust
  stochastic aggregation methods for distributed learning from heterogeneous
  datasets,'' in \emph{Proceedings of the AAAI Conference on Artificial
  Intelligence}, vol.~33, no.~01, Jul. 2019, pp. 1544--1551.

\bibitem{StochasticGradientDecent}
L.~Bottou, ``Large-scale machine learning with stochastic gradient descent,''
  in \emph{Proceedings of COMPSTAT'2010}.\hskip 1em plus 0.5em minus
  0.4em\relax Paris, France: Springer, August 2010, pp. 177--186.

\bibitem{DataSource}
\BIBentryALTinterwordspacing
M.~S. J. B.~D. Trifonov, ``Battery and heating data in real driving cycles,''
  2020. [Online]. Available: \url{https://dx.doi.org/10.21227/6jr9-5235}
\BIBentrySTDinterwordspacing

\bibitem{ARXmodel}
F.~A. {Ruslan}, K.~{Haron}, A.~M. {Samad}, and R.~{Adnan}, ``Multiple input
  single output (miso) arx and armax model of flood prediction system: Case
  study pahang,'' in \emph{IEEE 13th International Colloquium on Signal
  Processing its Applications}, Batu Ferringhi, Malaysia, 2017, pp. 179--184.

\bibitem{AdaptiveLearningFeature2}
W.~He, J.~T. Kwok, J.~Zhu, and Y.~Liu, ``A note on the unification of adaptive
  online learning,'' \emph{{IEEE} Trans. Neural Networks Learn. Syst.},
  vol.~28, no.~5, pp. 1178--1191, 2017.

\bibitem{ModelFitIndex}
R.~J.~C. de~Godoy and C.~Garcia, ``Optimal order selection for high order {ARX}
  models,'' \emph{Digit. Signal Process.}, vol. 108, p. 102897, 2021.

\bibitem{eRout9}
L.~{Thibault}, G.~{De Nunzio}, and A.~{Sciarretta}, ``A unified approach for
  electric vehicles range maximization via eco-routing, eco-driving, and energy
  consumption prediction,'' \emph{IEEE Transactions on Intelligent Vehicles},
  vol.~3, no.~4, pp. 463--475, 2018.

\end{thebibliography}

\end{document}